\documentclass[preprint,pra,aps,secnumarabic,eqsecnum,superscriptaddress]{revtex4}

\newcommand{\be}{\begin{equation}}
\newcommand{\beq}{\begin{equation}}
\newcommand{\ee}{\end{equation}}
\newcommand{\bea}{\begin{eqnarray}}
\newcommand{\eea}{\end{eqnarray}}
\newcommand{\ba}{\begin{array}}
\newcommand{\ea}{\end{array}}
\newcommand{\Id}[1] {\int \! \! \,{\rm d}^2 #1\:}

\renewcommand{\vr} {{\bf r}}

\newcommand{\vs} {{\bf s}}

\newcommand{\vj} {{\bf j}}

\begin{document}

\title{Semi-local density functional for the exchange-correlation
  energy of electrons in two dimensions}
 
\author{E. R\"as\"anen}
\email{erasanen@jyu.fi}
\preprint{submitted to Int. J. Quant. Chem.}
\affiliation{Nanoscience Center, Department of Physics, 
University of Jyv{\"a}skyl{\"a}, FIN-40014 Jyv{\"a}skyl{\"a}, Finland}

\author{S. Pittalis}
\affiliation{Department of Physics and Astronomy, University of Missouri,
Columbia, Missouri 65211, USA}

\author{J.\,G. Vilhena}
\affiliation{Laboratoire de Physique de la Mati\`{e}re
Condens\'{e} et Nanostructures,
Universit\'{e} Lyon I, CNRS, UMR 5586, Domaine scientifique de la
Doua, F-69622 Villeurbanne Cedex, France}
\affiliation{European Theoretical Spectroscopy Facility (ETSF)}

\author{M.\,A.\,L. Marques}
\affiliation{Laboratoire de Physique de la Mati\`{e}re
Condens\'{e} et Nanostructures,
Universit\'{e} Lyon I, CNRS, UMR 5586, Domaine scientifique de la
Doua, F-69622 Villeurbanne Cedex, France}
\affiliation{European Theoretical Spectroscopy Facility (ETSF)}

\date{\today}
 
\begin{abstract}
We present a practical and accurate density functional for the
exchange-correlation energy of electrons in two dimensions.
The exchange part is based on a recent two-dimensional 
generalized-gradient approximation derived by considering the limits 
of small and large density gradients. The fully local correlation part 
is constructed following the Colle-Salvetti 
scheme and a Gaussian approximation for the pair density. 
The combination of these expressions is shown to provide an efficient
density functional to calculate the total energies of two-dimensional
electron systems such as semiconductor quantum dots. Excellent
performance of the functional with respect to numerically exact 
reference data for quantum dots is demonstrated.
\end{abstract}

\maketitle
 
\section{Introduction}

The practical capability of density-functional theory~\cite{dft} 
(DFT) to capture many-particle properties of physically
and chemically relevant systems crucially depends on the availability of 
good approximations for the exchange-correlation energy functional.
Significant advances have been achieved by 
means of, e.g., local and semi-local approximations, orbital
functionals, and hybrid functionals~\cite{functionals}.
Most of these constructions have been carried out considering 
three spatial dimensions (3D) explicitly, which is natural in view
of atoms, molecules, and solids. Consequently, however,
the field of low-dimensional physics has been left, for the most part,
outside the coverage of DFT. For example, density functionals
developed particularly for 3D fail when applied to 
(quasi-)two-dimensional (2D) systems~\cite{kim,pollack,perdew}.

At present, 2D nanoscale structures have a rich variety including, e.g.,
quantum Hall bars, semiconductor quantum dots~\cite{qd} and rings, and
metal-oxide-semiconductor-based field-effect transistors~\cite{vignalebook}. 
It should be noted that the building block here is the 
{\em quasi-2D electron gas}, which can be treated, however, by
a ``pure'' 2D approach, i.e., on a 2D grid with Coulombic electron-electron
interactions. More explicitly,
the degrees of freedom along the third dimension have been suppressed
and the system is effectively 2D. Then, the influence of the 
surrounding host material is taken into account with the effective-mass 
approximation manifesting itself as an effective
mass and a dielectric constant in the 2D Hamiltonian.

Within DFT, 2D systems are usually dealt with the 2D local-density
approximation~\cite{rajagopal,attaccalite} 
(LDA), which is accurate in terms of 
{\em total} energies in many cases~\cite{nicole,saarikoski}. 
Problems are expected to arise in highly inhomogeneous systems, 
or when considering the very strong interaction regime~\cite{paola},
or close to dimensional crossovers~\cite{x2}. 
Developments in 2D beyond
the LDA have been carried out recently, leading to accurate density 
functionals for both exchange~\cite{x1,x2,gga,gamma} and 
correlation~\cite{c1,2dcs}. 
Good results have been obtained also from orbital 
functionals designed to follow the dimensional 
crossover~\cite{perdew}, simple orbital-free total-energy 
functionals~\cite{simple}, as well as 
from DFT specialized for strongly interacting electrons~\cite{paola}.

In this work we present a 2D density functional 
combining the 2D generalized-gradient approximation~\cite{gga}
(GGA) for the exchange and a local density functional for the 
correlation~\cite{2dcs}. The functional is shown to be accurate not only for 
the exchange and correlation energies, respectively, but also for 
the total energy. Thus, it could be
readily used for the calculation of, e.g., chemical potentials and
addition energies of quantum-dot systems -- quantities directly
available from Coulomb-blockade transport
experiments~\cite{blockade}. As a numerically efficient method,
the functional is also a promising tool for numerical studies on
confined quantum Hall systems typically dealt with single-particle
theories or with the Thomas-Fermi approximation~\cite{afif}.

\section{Theory}

We focus here on the Kohn-Sham (KS) scheme within DFT 
to obtain the ground-state total energies $E_{\rm tot}$ 
and densities $n(\vr)$ of a 2D system
containing $N$ Coulomb-interacting electrons. The total
energy can be written, in Hartree atomic units, as a density
functional
\be
  \label{etot}
  E[n] = T_s [n] + 
  E_{\rm H}[n] + E_{\rm xc}[n] 
  + \Id{r} \,v(\vr)
  n(\vr), 
\ee
where $T_s[n]$ is the KS
kinetic energy functional, $E_{\rm H}[n]$ is
the Hartree energy, $v(\vr)$ is the external 
scalar potential, and
$E_{\rm xc}[n]$ is the exchange-correlation
energy functional. 
The latter can be decomposed into the
exchange and correlation parts as 
$E_{\rm xc} [n]= E_x[n] + E_c[n]$. In the following 
sections we consider approximations for these parts separately. 

It should be noted that for $E_x[n]$ we use an approximation which 
depends on the spin density and thus also applies to spin-polarized states
within the spin-DFT formalism,
whereas for $E_c[n]$ the approximation depends only the total density, and 
its derivation suggests that is suited for
spin-unpolarized systems. This restriction for the correlation calls 
for a spin-dependent extension.

\subsection{Exchange}

The following 2D-GGA is based on the idea of Becke~\cite{becke2,becke3}
for 3D systems, which was extended to 2D in Ref.~\cite{gga}.
The spin-density ($n_\sigma$) functional for the exchange energy 
can be expressed as
\be\label{EXCA}
  E_x [n_\sigma] = - \pi \sum_{\sigma=\uparrow,\downarrow}
  \Id{r} n_{\sigma}(\vr)
  \int\!\!{\rm d} s \: \bar{h}_{\rm{x},\sigma}(\vr, s),
\ee
where $\bar{h}_{x,\sigma}(\vr, s)$ is the {\em cylindrical average}
with respect to  $\vs=\vr'-\vr$ (with $s=|\vr'-\vr|$)
of the exchange-hole (or Fermi-hole) function $h_{x,\sigma}(\vr,\vr')$
around the reference point $\vr$. 
Note that here
we use a definition of the exchange energy for which the
corresponding hole function is positively defined. 
The Taylor expansion of  $\bar{h}_{x,\sigma}(\vr, s)$ 
with respect to $s$ up to the second order gives 
\begin{equation}
  \label{taylor2}
  \bar{h}_{x,\sigma}(\vr,s)  =  n_{\sigma}(\vr) + C^{\sigma}_x(\vr)s^2  + \ldots ,
\end{equation}
where 
\begin{equation}
  \label{C}
  C^{\sigma}_x(\vr) = \frac{1}{4}\left[ \nabla^2 n_{\sigma}(\vr) -2\tau_\sigma(\vr)
  + \frac{1}{2}\frac{\left| \nabla n_\sigma(\vr) \right|^2}{n_\sigma}
  + 2 \frac{\vj^2_{p,\sigma}(\vr)}{n_\sigma(\vr)} \right]
\end{equation}
is the local curvature of the
exchange hole around the given 
reference point $\vr$~\cite{becke2,Dobson:93,x1,2delf}.
Here $\tau_\sigma$ is twice the kinetic-energy density and 
$\vj_{p,\sigma}$ is the paramagnetic current density.

In the small density-gradient limit (SGL), we take the
homogeneous 2D electron gas (2DEG) as the reference system.
When the inhomogeneity is small, we may write
\be
\label{modela}
\bar{h}_{x,\sigma}(\vr,s) 
= \left\{ \begin{array}{ll}
\left[1+ a_{\sigma}(\vr)s^2 + b_{\sigma}(\vr)s^4 +
  \ldots \right]\bar{h}^{\rm 2DEG}_{x,\sigma}(s), & \;\; \textrm{if $k_{F,\sigma} s < z$}\\
\bar{h}^{\rm 2DEG}_{x,\sigma}(s), & \;\; \textrm{if $k_{F,\sigma} s \geq z$}.
\end{array} \right.
\ee
Here $\bar{h}^{\rm 2DEG}_{x,\sigma}$ is the exact exchange-hole
function for the 2DEG~\cite{gori}, $k_{F,\sigma}=\sqrt{4\pi
  n_{\sigma}}$ is the 2D Fermi momentum, and 
where $z$ is the first zero of $J_1$, i.e., the Bessel function of the 
first kind appearing in $\bar{h}^{\rm 2DEG}_{x,\sigma}$.
Comparing Eqs.~(\ref{taylor2}) and (\ref{modela}), using
the 2D Thomas-Fermi expression for $\tau_{\sigma}$, and enforcing
the exact normalization of $\bar{h}_{x,\sigma}$ lead to expressions
for $a_{\sigma}$ and $b_{\sigma}$ (see Ref.~\cite{gga} for details),
and finally the SGL can be written as
\be\label{SGL}
  E^{\rm SGL}_{x,\sigma} 
  = -\frac{5}{48\sqrt{\pi}}\left[  \frac{I(0)I(3)-I(1)I(2)}{I(3)} \right]
\Id{r} \frac{|\nabla n_{\sigma}(\vr)|^2}{n^{3/2}_{\sigma}(\vr)},
\ee
where 
\be
  I(m) = \int_{0}^{z}\!\! {\rm d}y\: y^m J^2_1(y).
\ee

In the large density-gradient limit (LGL) 
the {\em density gradient} dominates over the other terms
in $\bar{h}_{x,\sigma}$, and, secondly, the exchange hole vanishes
at large $s$ following a decay function denoted here as $F$. Thus, we may write
\be
  \label{lg2}
  \bar{h}_{x,\sigma}(\vr,s) \approx
  \left[\frac{1}{8} \frac{\left|\nabla n_\sigma(\vr) \right|^2}{n_\sigma(\vr)} s^2 \right]
  F(\alpha_{\sigma}(\vr)\, s),
\ee
where $F(y) = e^{-y^2}$ corresponds to a Gaussian approximation for
the exchange hole. The parameter $\alpha_{\sigma}$ can be determined by
enforcing again the exact normalization constraint for $\bar{h}_{x,\sigma}$. 
The
resulting expression for the LGL of the exchange energy in Eq.~(\ref{EXCA})
is
\be\label{LGL}
  E^{\rm LGL}_{x,\sigma} 
  = -\frac{\pi^{1/4}}{2^{3/2}}G(2)G^{-3/4}(3)
\Id{r} \frac{|\nabla n_{\sigma}(\vr)|^{1/2}}{n^{3/4}_{\sigma}(\vr)},
\ee
where 
\be
  G(m) = \int_{0}^{\infty}\!\!{\rm d} y\: y^m e^{-y^2}.
\ee

Finally we interpolate the results for the SGL [Eq.~(\ref{SGL})] and
LGL [Eq.~(\ref{LGL})] leading to the 2D-GGA expression for the
exchange energy,
\be
\label{2dgga}
 E^{\rm GGA}_x[n_{\sigma},\nabla n_{\sigma}] =
E^{\rm LDA}_x[n_{\sigma}] 
-\zeta\sum_{\sigma=\uparrow,\downarrow}
\Id{r} \frac{|\nabla n_\sigma(\vr)|^2}{n^{3/2}_{\sigma}(\vr)\left[1+\gamma\frac{|\nabla n_\sigma(\vr)|^2}{n^3_{\sigma}(\vr)}\right]^{3/4}}\,,
\ee
where 
\be
E^{\rm LDA}_x[n_{\sigma}] = -\frac{8}{3\sqrt{\pi}}\Id{r} n^{3/2}_{\sigma}(\vr)
\ee
is the 2D-LDA exchange energy. The parameters $\zeta$ and $\gamma$ are
fitted to a set of parabolic quantum dots yielding 
$\zeta=0.003317$ and $\gamma=0.008323$~\cite{gga}. 
Interestingly, these values are close to the parameters 
found by Becke by fitting a series of noble-gas atoms~\cite{becke3}.

\subsection{Correlation}

Here we review the derivation of the correlation energy functional
presented in Ref.~\cite{2dcs}.
The correlation energy can be expressed as~\cite{CS1,CS2,2dcs}
\be\label{ce2}
  E_c = \Id{r}\Id{s}
  n_{2,{\rm SD}}(\vr,\vs) \frac{\varphi^2(\vr,\vs)-2\varphi(\vr,\vs)}{s}
  \,,
\ee
where $n_{2,{\rm SD}}(\vr,\vs)$ is the pair density calculated 
using a single Slater determinant (SD) generated within DFT
from the occupied KS orbitals [see Eqs.~(\ref{cs}) and (\ref{KS}) below].
Further, the quantity
\be\label{cf}
  \varphi(\vr,\vs) = \left[ 1 - \Phi(\vr)(1 + \alpha s) \right]e^{-\beta^2(\vr) s^2}
  \,
\ee
describes the correlation between electron pairs. Here
$\vs=\vr_1-\vr_2$ and $\vr=(\vr_1+\vr_2)/2$ represent the
relative and center-of-mass coordinates of a representative electron pair, 
respectively. The quantities $\alpha$, $\beta$, and $\Phi$ 
act as correlation factors to be determined as follows.

\begin{itemize}

\item First, $\alpha$ can be found by considering
the cusp condition for a 2D singlet many-body wavefunction.
This corresponds to a situation when two electrons are brought to 
the same point. Application of the exact result of Rajagopal 
{\em et al.}~\cite{RJK} to the model wave function
\be\label{cs}
  \Psi(\vr_1 \sigma_1,..., \vr_N \sigma_N) =
  \Psi_{\rm SD}(\vr_1 \sigma_1,..., \vr_N \sigma_N) \prod_{i < j} \left[1-\varphi(\vr_i,\vr_j) \right]
  \,
\ee 
-- from which 
the correlation energy given in Eq.~(\ref{ce2}) is obtained ~\cite{2dcs} --
yields $\alpha=1$.

\item Second, we introduce $\beta$ as a {\em local} quantity
determining the local correlation length. We can estimate this length
by comparing the "correlation area", i.e., area integral over the
exponential part of Eq.~(\ref{cf}), to the area of a circle enclosing,
on the average, one electron, i.e., $\pi r_s^2$, where 
$r_s(\vr)=1/\sqrt{\pi n(\vr)}$. The comparison yields 
$\beta(\vr)=q\sqrt{n(\vr)}$, where we introduce $q$ as a fitting parameter.

\item Third, the integral of $\varphi(\vr,\vs)$ [Eq.~(\ref{cf})]
over the relative coordinate $\vs$ must vanish
(see Refs. ~\cite{CS2,MS} for details). 
This leads to a relation 
$\Phi(\vr) = \beta(\vr)/\left[\beta(\vr) + \sqrt{\pi}/2\right]$.

\end{itemize}

To further simplify the expression for the correlation energy
[Eq.~(\ref{ce2})], we use a Gaussian approximation~\cite{MS,Parr88} 
for the SD pair density,
\be
  n_{2,{\rm SD}}(\vr,\vs) = n_{2,{\rm SD}}(\vr) 
  e^{-s^2/\gamma^2(\vr)}
  \,.
\ee
Applying the exact sum rule on the pair density yields
\be
  n_{{\rm SD}}(\vr) =  \frac{2}{N-1}\int\!\!d^2s\;n_{2,{\rm
      SD}}(\vr,\vs) =  \frac{2\pi}{N-1}\,n_{2,{\rm SD}}(\vr)\gamma^2(\vr)
  \,,
\ee
where we can use the well-known relation in the SD case:
$n_{2,{\rm SD}}(\vr)=\frac{1}{4}n_{{\rm SD}}^2(\vr)$. Further,
we can associate the SD density $n_{{\rm SD}}$ with the density
in the second item of the list above. Taken together, and performing
the integration over $s$ in Eq.~(\ref{ce2}), leads to the correlation 
energy
\be\label{Ec}
  E_{\rm c}[n] = \int\!\!d^2r\; n(\vr)\,\epsilon_{\rm c}(\vr)
  \,,
\ee
where $\epsilon_{\rm c}(\vr)$ is the local correlation energy 
per electron having an expression
\bea\label{epsilon}
  \epsilon_{\rm c}(\vr) = \frac{\pi}{2q^2}\Bigg\{
    \frac{\sqrt{\pi}\:\beta(\vr)}{2\sqrt{2+c}}[\Phi(\vr) - 1]
   +\frac{\Phi(\vr)[\Phi(\vr)-1]}{2+c}
  \nonumber\\   + \frac{\sqrt{\pi}\:\Phi^2(\vr)}{4\beta(\vr)(2+c)^{3/2}}
   + \frac{\sqrt{\pi}\:\beta(\vr)}{\sqrt{1+c}}[\Phi(\vr)-1]
   + \frac{\Phi(\vr)}{1+c}
  \Bigg\}\,,
\eea
with $c = \pi/\left[2(N-1)q^2\right]$. 
We point out that this expression includes an {\em ad-hoc} 
modification in the first term, 
$[\Phi(\vr) - 1]^2 \rightarrow \left[\Phi(\vr) - 1\right]$. 
This modification is introduced to better reproduce 
the reference values of numerically accurate correlation energies 
for a set of quantum dots and the 2DEG~\cite{2dcs}. 

Equation~(\ref{Ec}) with Eq.~(\ref{epsilon})
define an explicit density functional for the correlation
energy. It is self-interaction free (in the sense that it is equal to zero for
one particle systems, $N=1$)
and depends on a single fitting parameter $q$ (see below).
We point out that the functional is not size-consistent due
to the nonlinear dependence on $N$. 
In practice, however, this does not affect the performance of the 
approximation when considering finite 2D systems with fixed $N$,
e.g., semiconductor quantum dots.

The fitting parameter $q=3.9274$ is chosen to
reproduce the {\em exact} correlation energy $E_{\rm c} =
E_{\rm tot}-E^{\rm EXX}_{\rm tot}$ for the singlet state of a
two-electron parabolic quantum dot (2D harmonic oscillator) with
the confining strength $\omega=1$. Here $E_{\rm tot}=3$ is Taut's
analytic result~\cite{taut} and $E^{\rm EXX}_{\rm tot}$ is the
exact-exchange result. For consistency, and in order to provide
a predictive approximation, the fitting parameter $q$ is 
then kept fixed for {\em all} systems.

\section{Results}

 The sum of Eqs.~(\ref{2dgga}) and 
(\ref{Ec}) is the approximation we employ for the $E_{xc}$.
We test this functional in the calculation of ground-state
total energies of spin-unpolarized, closed-shell quantum dots.
Thus $n_{\uparrow}(\vr)=n_{\downarrow}(\vr) =n(\vr)/2$ and, as a consequence,
we can restrict our calculations to the standard DFT scheme.
The total energy is then obtained from Eq. (\ref{etot}) in connection with
solving self-consistently the KS equation
\be\label{KS}
\left[-\frac{1}{2}\nabla^2 + v_{\rm KS}(\vr)\right]\varphi_i(\vr)=\epsilon_i\varphi_i(\vr).
\ee
The KS potential is given by
$v_{\rm KS}(\vr)=v_{\rm H}(\vr)+v_{xc}(\vr)+v(\vr)$, 
where
the Hartree potential is computed as
\be
v_{\rm H}(\vr) = \Id r' \frac{n(\vr')}{|\vr - \vr'|},
\ee
the exchange-correlation potential is obtained from
\be
v_{xc}(\vr) = \frac{\delta E_{xc}}{\delta n(\vr)},
\ee
and $v(\vr)$ is the given external potential. 
The KS orbitals $\varphi_i(\vr)$ provide the
the total density as $n(\vr)=\sum_{i=1}^N |\varphi_i(\vr)|^2$
($N/2$ electrons for each spin-channel), and
$\epsilon_i$ are the corresponding KS energies. 

Below, we compare the reference results available from (numerically)
exact calculations to self-consistent DFT
results obtained using the real-space {\tt octopus} 
code~\cite{octopus} with the proposed functional and with the LDA,
respectively.

\subsection{Parabolic quantum dots}

First we consider a set of 2D parabolic (harmonic)
quantum dots, where the external confining potential in 
Eq.~(\ref{etot}) is given by $v(r)=\omega^2 r^2/2$.
Table~\ref{table_qd}
\begin{table}
  \caption{\label{table_qd} Total energies (in
  atomic units) for parabolic quantum dots. We compare numerically
exact results ($E_{\rm tot}^{\rm ref}$) to our functional ($E_{\rm tot}^{\rm here}$)
and to the local-density approximation ($E_{\rm tot}^{\rm LDA}$).
The last row contains the mean percentage error $\Delta$.
}
  \begin{tabular}{c c c c c}
  \hline
  \hline
  $\quad N \quad$ & $\omega$ & $E_{\rm tot}^{\rm ref}$ &  $E_{\rm tot}^{\rm here}$
  &  $E_{\rm tot}^{\rm LDA}$ \\
  \hline
  2  & 1           & $3^*$             & 3.026 & 3.066 \\
  2  & 1/4         & $0.9324^\dagger$  & 0.936 & 0.952 \\
  2  & 1/6         & $2/3^*$           & 0.668 & 0.682 \\
  2  & 1/16        & $0.3031^\dagger$  & 0.300 & 0.308 \\
  6  & $1/1.89^2$  & $7.6001^\ddagger$ & 7.629 & 7.632 \\
  6  & 1/4         & $6.995^\dagger$   & 7.009 & 7.012 \\
  6  & 1/16        & $2.528^\dagger$   & 2.528 & 2.534 \\
  12 & $1/1.89^2$  & $25.636^\ddagger$ & 25.72 & 25.67 \\
  \hline
  \multicolumn{3}{l}{$\Delta$}                                    & 0.42\%  & 1.2\% \\
  \hline
  \hline
  \end{tabular}
  \begin{flushleft}
    $^*$   Analytic solution by Taut from Ref.~\onlinecite{taut}.
  $^\dagger$  CI data from Ref.~\onlinecite{rontani}.
  $^\ddagger$ Diffusion QMC data from Ref.~\onlinecite{pederiva}.
  \end{flushleft} 
\end{table}
shows the total energies for $N=2\ldots 12$ with various confinement
strengths $\omega$. The reference data $E_{\rm tot}^{\rm ref}$ have been
collected from analytic results by Taut~\cite{taut}, 
configuration-interaction calculations by Rontani {\em et al.}~\cite{rontani},
and diffusion quantum Monte Carlo (QMC) calculations by Pederiva {\em et al.}~\cite{pederiva} Overall, both LDA and our functional perform very well with 
respect to the reference data, the mean percentage errors being
$\sim 1.2\,\%$ and $0.42\,\%$, respectively. However, in view of the excellent
performance of the LDA in terms of total energies, it is remarkable 
that the present functional reduces the error further by a factor of three.
Moreover, it should be noted that the LDA total
energy has the well-known error compensation from exchange and correlation 
energies, respectively~\cite{nicole}. In the results shown in 
Table~\ref{table_qd}, for example, the LDA overestimates (underestimates) 
the exchange (correlation) energy by $10\ldots 20\,\%$, whereas the 
corresponding errors in our functional are significantly 
smaller~\cite{gga,2dcs}.

As seen in Table~\ref{table_qd}, the total energy of the largest system with 
$N=12$ is obtained by the LDA more accurately than by our functional.
This raises a question whether the LDA would considerably outperform our
functional in the important large-$N$ limit. Therefore, we also tested
the filling-factor $\nu=2$ state of a $N=48$ quantum dot, for which relatively
accurate variational QMC data is available (see Fig. 8 in Ref.~\cite{droplet}). 
Remarkably, the relative errors of the present functional and the LDA are 
only $\sim 0.1\,\%$ (note that in Ref.~\cite{droplet} a different DFT code
was used producing a slightly larger error in the LDA). This test confirms
that the present functional is valid also in relatively large 2D systems.
This is expected in view of the good reproducibility of the 2DEG result
imposed on both the exchange [see Eq.~(\ref{modela})] and correlation
(see Fig. 1 in Ref.~\cite{2dcs}).

\subsection{Rectangular quantum dots}

Next we consider the total energies of 
rectangular quantum dots with a side-length ratio $\chi$ and 
a total area of $\pi^2$ (a.u.) enclosed by hard-wall boundaries. 
As the reference results we use the variational
QMC data in Ref.~\cite{rectapaper}, where the
choice for the dot area was motivated by a rational-valued
single-electron energy spectrum, $E_{ij}=(\chi\,i+j/\chi)/2$, where
$(i,j)=1,2,3,\ldots$. Moreover, the dot size corresponds to an
realistic area of $\sim 900\;{\rm nm}^2$, when using the effective-mass 
approximation for electrons in GaAs~\cite{rectapaper}.

The total-energy results are shown in Table~\ref{table_qd_recta}.
\begin{table}
  \caption{\label{table_qd_recta} Total energies (in
  atomic units) for rectangular quantum dots with a total area of $\pi^2$ and 
side-length ratio $\chi$.
We compare numerically accurate quantum Monte Carlo results~\cite{rectapaper}
($E_{\rm tot}^{\rm ref}$) to our functional ($E_{\rm tot}^{\rm here}$)
and to the local-density approximation ($E_{\rm tot}^{\rm LDA}$).
The last row contains the mean percentage error $\Delta$.
}
  \begin{tabular}{c c c c c}
  \hline
  \hline
$\quad \chi \quad$  & $\quad N \quad$ & $E^{\rm ref}_{\rm tot}$ & $E_{\rm tot}^{\rm here}$ & $E_{\rm tot}^{\rm LDA}$ \\
\hline
$1$ &  2 & 3.273   & 3.312  & 3.357 \\
 &  6 & 26.97  & 26.98  & 27.10 \\
 &  8 & 46.79  & 46.86  & 46.99 \\
 & 12 & 103.34 & 103.1  & 103.2 \\
 & 16 & 178.50 & 178.4  & 178.5 \\
  \hline
$2$ &  2 & 3.696 & 3.674 & 3.735 \\
 &  4 & 12.32  & 12.36 & 12.45 \\
 &  6 & 27.15  & 27.25 & 27.36 \\
 &  8 & 47.82  & 47.69 & 47.80 \\
 & 12 & 102.26 & 102.1 & 102.2 \\
 & 16 & 177.80 & 178.0 & 178.1 \\
  \hline
$3$ &  2 & 4.375 & 4.321 & 4.403 \\
 &  4 & 12.99 & 12.95 & 13.08 \\
 &  6 & 26.69 & 26.75 & 26.91 \\
 &  8 & 46.35 & 46.49 & 46.67 \\
 & 12 & 103.46 & 103.4 & 103.5 \\
 & 16 & 177.37 & 177.1 & 177.3 \\
  \hline
  \multicolumn{3}{l}{$\Delta$} & $0.34\%$  & $0.57\%$ \\
  \hline
  \hline
  \end{tabular}
\end{table}
Again, the present functional is more accurate than 
the LDA: in this case the LDA error is reduced by 
a factor of $1.7$. There is no clear tendency in the
accuracies as a function of $N$. Precise determination of
the $N$-dependence would require highly accurate
numerical calculations and reference data. In this respect
the present boundary conditions (hard walls) are
problematic due to the huge number of grid points needed.

\section{Summary}

We have considered a practical and accurate density functional for 
the energy of electrons confined in two dimensions.
The exchange contribution is a semi-local generalized-gradient 
approximation, and the correlation part has been built on the
Colle-Salvetti scheme with a Gaussian approximation
for the pair density.  We have verified that self-consistent application 
performs very well for a variety of two-dimensional quantum-dot systems.
On the average, the present functional was found to considerably reduce the
error in total energies over the local-density approximation.
Preliminary tests suggest also good performance for large electron 
numbers with a modest computational cost. Future developments
include an extension of the correlation part to spin-polarized 
systems.

\begin{acknowledgments}
We thank Ari Harju for the variational quantum Monte Carlo data.
This work was supported by the Academy of Finland and 
the EU's Sixth Framework Programme through the ETSF e-I3.
M.A.L. Marques acknowledges partial support by the French ANR program
(ANR-08-CEXC8-008-01) and the Portuguese FCT (PTDC/FIS/73578/2006).
J.G.V. acknowledges financial support by the Portuguese FCT through
Project No. SFRH/BD/38340/2007. Computer resources were provided by the 
LCA of the University of Coimbra, GENCI (x2009096017), and Freie 
Universit\"at Berlin.
S. Pittalis acknowledges support by DOE grant DE-FG02-05ER46203 and 
Prof. Giovanni Vignale.
\end{acknowledgments}

%
%
 
\end{document}